\documentclass[12pt,preprint]{aastex}
\usepackage{graphicx}        
\usepackage{natbib}         
\usepackage{amssymb}        
\usepackage{amsmath}
\usepackage{color}           
\usepackage{url}             
\usepackage{epstopdf}

\renewcommand{\vec}[1]{ {\mathbf #1} }

\shorttitle{2D or not 2D}
\shortauthors{Garaud and Brummell}

\begin{document}



\title{2D or not 2D: the effect of dimensionality on the dynamics of fingering convection at low Prandtl number }
\author{Pascale Garaud$^1$ and Nicholas Brummell$^2$}
\affil{$^1$ Department of Applied Mathematics and Statistics, Baskin School of Engineering, University of California Santa Cruz, 1156 High Street, Santa Cruz CA 95060}


\begin{abstract}
Fingering convection (otherwise known as thermohaline convection) is an instability that occurs in stellar radiative interiors in the presence of unstable compositional gradients. Numerical simulations have been used in order to estimate the efficiency of mixing induced by this instability. However, fully three-dimensional (3D) computations in the parameter regime appropriate for stellar astrophysics (i.e. low Prandtl number) are prohibitively expensive. This raises the question of whether two-dimensional (2D) simulations could be used instead to achieve the same goals. In this work, we address this issue by comparing the outcome of 2D and 3D simulations of fingering convection at low Prandtl number. We find that 2D simulations are never appropriate. However, we also find that the required 3D computational domain does not have to be very wide: the third dimension need only contain a minimum of two wavelengths of the fastest-growing linearly unstable mode to capture the essentially 3D dynamics of small-scale fingering. Narrow domains, however, should still be used with caution since they could limit the subsequent development of any large-scale dynamics typically associated with fingering convection.
\end{abstract}

\maketitle
\section{Introduction}
\label{sec:intro} 

Fingering convection (also called thermohaline convection) was first discussed in the astrophysical context by \citet{Ulrich1972} \citep[see also][]{Ulrich1971}. The basic fingering instability, which occurs in systems that are compositionally unstably stratified and thermally stably stratified, is common in stars whose surface is polluted by high mean molecular weight material. This can happen through the accretion of material from infalling planets or debris disks \citep{vauclair2004mfa,garaud11,Dealal2013} or from more evolved companion stars \citep{stancliffe2007cem,Angelou2012}. Fingering convection can also result from the radiative levitation of heavy elements such as iron from deeper regions upwards in relatively high-mass stars \citep{theado09,Zemskovaal14}, or from off-center nuclear burning in post main sequence stars \citep{CharbonnelZahn07,denissenkov2010,DenissenkovMerryfield2011}.

In the typical conditions encountered in stellar interiors, instability to fingering convection only depends on the value of the density ratio $R_0$, defined as 
\begin{equation}
R_0 = \frac{\alpha \left( \frac{dT_0}{dr} - \frac{dT_{\rm ad}}{dr} \right) }{\beta \frac{d\mu_0}{dr} } \mbox{   ,   } 
\end{equation}
where $T_0(r)$ is the local temperature profile in the stellar region considered, $T_{\rm ad}(r)$ is the temperature profile that region would have were it to be adiabatically stratified, $\mu_0(r)$ is the local mean molecular weight profile, and $\alpha$ and $\beta$ are derivatives of the equation of state defined as
\begin{equation}
\alpha = - \frac{1}{\rho} \left(\frac{\partial \rho}{\partial T} \right)_{P,\mu} \mbox{   ,   } \beta = \frac{1}{\rho}\left( \frac{\partial \rho}{\partial \mu}\right)_{P,T}  \mbox{   .  } 
\end{equation}
The density ratio is therefore the ratio of the density gradient due to the stabilizing thermal stratification to the density gradient due to the destabilizing compositional stratification.  
 Note that $R_0$ is also commonly written as 
\begin{equation}
R_0 = \frac{\delta  (\nabla - \nabla_{\rm ad})}{\phi \nabla_\mu} \,
\end{equation}
where $\nabla$, $\nabla_{\rm ad}$ and $\nabla_{\mu}$ have their usual astrophysical definitions, and 
\begin{equation}
\delta = - \left( \frac{\partial \ln \rho}{\partial \ln T} \right)_{P,\mu} =  \alpha T \mbox{  and  } \phi =  \left( \frac{\partial \ln \rho}{\partial \ln \mu} \right)_{P,T}  = \beta \mu  \mbox{   .  } 
\end{equation} 

As shown by \citet{bainesgill1969}, linear instability to fingering convection only occurs when 
\begin{equation}
1 < R_0 < R_{\rm crit} = \tau^{-1} = \frac{\kappa_T}{\kappa_\mu} , 
\end{equation}
where $\kappa_T$ and $\kappa_\mu$ are the thermal and compositional diffusivities respectively.  The lower limit ($R_0 = 1$) corresponds to the onset of standard overturning convection, while the upper limit ($R_0 = R_{\rm crit}$) corresponds to the critical point of marginal stability to fingering convection.  As shown by \citet{Medranoal15}, $\tau$ varies between $10^{-9}$ and $10^{-5}$ in the interiors of main sequence stars, and increases only up to about $10^{-3}$ in degenerate regions of more evolved stars (e.g. red giants or white dwarfs). This shows that $R_{\rm crit}$ is always much greater than one, so that stellar interiors can very easily be destabilized: even a tiny adverse compositional gradient can trigger fingering instabilities. As a result, fingering convection is very common, and should be taken into account when considering mixing in stellar evolution models.   

An important difference in the behavior of fingering instabilities in astrophysical and geophysical systems comes from the value of the Prandtl number (${\rm Pr} = \nu/\kappa_T$, where $\nu$ is the kinematic viscosity): this parameter is usually larger than one in most geophysical fluids, but is asymptotically small \citep[typically, at most one order of magnitude larger than $\tau$; see][]{Medranoal15} in stellar interiors. As a result, fingering structures which are typically fairly laminar in the geophysical context are instead very turbulent in the astrophysical one. This complicates the study of the saturation of the instability, which is necessary to estimate the rate of heat and compositional transport induced by fingering convection. 

In the past few years, thanks to the development of fast parallel algorithms and advances in supercomputing, it has become possible to follow the nonlinear development of fingering instabilities in three dimensional (3D) simulations at relatively low diffusivity ratio and low Prandtl numbers, albeit not yet at actual stellar values of these parameters. The first of such studies were by \citet{denissenkov2010} using 2D simulations and by \citet{Traxleral2011} and \citet{Brownal2013} in 3D. The 3D simulations covered a wide range of parameter space with ${\rm Pr}$ and $\tau$ varying from 0.3 down to 0.01, and the density ratio was varied across the instability range (from 1 to $\tau^{-1}$). Using them, \citet{Brownal2013} were able to propose an analytical model for turbulent transport by fingering convection at low ${\rm Pr}$ and $\tau$. This model has only one free parameter, which can be calibrated using the 3D simulations. 
Once calibrated, the \citet{Brownal2013} model correctly predicts the transport rate for both heat and composition for {\it all} available simulations in which $\tau < {\rm Pr}$ within a factor of 2 at worst, often much better. 

Whether this model remains as accurate for $\Pr$ and $\tau$ lower than $10^{-2}$ still remains to be determined. However, cubic-domain 3D simulations at very low ${\rm Pr}$ and $\tau$ rapidly become computationally prohibitive since the resolution must  be increased as ${\rm Pr}$ and $\tau$ decrease, in order to fully resolve the various boundary layers that form at the edge of the fingers. In addition, and as discussed by \citet{Traxleral2011b} and \citet{Medranoal15}, secondary large-scale instabilities can develop spontaneously from fingering convection, and the latter substantially affect the turbulent transport properties of the system. To study them requires very large computational domains, containing at the very least 20 to 30 wavelengths of the fastest-growing fingering mode in both horizontal and vertical directions, and often much more than that. Again, using such large domains in 3D, together with low ${\rm Pr}$ and $\tau$, is currently computationally prohibitive. 

For both reasons, it is very tempting to go back to 2D simulations either for very low diffusivity studies, or for very large domain studies. In fact, 2D fingering convection simulations used to be common in the oceanographic literature only 20 years ago \citep[e.g.][]{shen1995,stern2001sfu}, and are still often in use today \citep{SimeonovStern2007,radko2008ddm,Sreenivasal09,SinghSrinivasan14} in that context. \citet{stern2001sfu} showed by comparing 2D and 3D simulations at ${\rm Pr} = 7$ that the typical turbulent fluxes estimated from 2D simulations do not vary from those obtained in 3D simulations by more than a factor of a few. \citet{DenissenkovMerryfield2011} claim to have run both 2D and 3D simulations of fingering convection for ${\rm Pr}$ and $\tau$ both $O(10^{-6})$, and to have found that the effective turbulent diffusivities of composition are very similar in the two cases. 

However, the work of \citet{Radko2010} casts strong doubts on the claims of \citet{DenissenkovMerryfield2011}. Indeed, he showed both analytically and numerically that the saturation of the fingering instability at low ${\rm Pr}$ in 2D can be attributed to the development of strong shear layers that destroy the fingers and completely change the overall behavior of the induced turbulence. A similar process was not seen in the 3D simulations of  \citet{RadkoSmith2012} and \citet{Brownal2013}. This raises a number of practical questions: (1) Is shear indeed present in low Prandtl number fingering \citep[as claimed by][]{Radko2010}, or not \citep[as implied by][]{DenissenkovMerryfield2011}? (2) More generally, under which circumstances, if any, can 2D simulations be used to study fingering convection at low Prandtl number? (3) If 2D simulations are not appropriate, then how thick does a domain have to be in order to approximate 3D results appropriately? In this work, we answer all three questions. 

Section \ref{sec:model} outlines our model equations, boundary conditions, and domain geometry. In Section \ref{sec:highR0}, we first compare 2D and 3D results in high density ratio simulations. In Section \ref{sec:lowR0}, we present a similar comparison but in low density ratio simulations. As we shall demonstrate, in both cases 2D and 3D results are fundamentally different, but for different reasons. Section \ref{sec:ccl} discusses our findings and proposes acceptable compromises in terms of running 3D fingering simulations as cheaply as possible. 

\section{Model}
\label{sec:model}

In what follows, we study the dynamics of homogeneous fingering convection in the Boussinesq approximation, as introduced for the purpose of numerical simulations by \citet{shen1995} \citep[see also][]{Radko2003mlf}. In this formalism, the background temperature and composition profiles are assumed to be linear, with locally constant gradients $dT_0/dr$ and $d\mu_0/dr$ (see Section \ref{sec:intro}), while all perturbations to that background (including the temperature $T$, the pressure $p$, the velocity field $\vec{u}=(u,v,w)$ and the mean molecular weight $\mu$) are assumed to be triply periodic in the domain considered. 

The non-dimensional governing equations are given by:
\begin{eqnarray}
&& \frac{1}{\rm{Pr}}\left(\frac{\partial \vec{u}}{\partial t} + \vec{u}\cdot\nabla \vec{u}\right) = -\nabla p + (T-\mu) \vec{e}_z + \nabla^2 \vec{u}, \nonumber \\
&& \frac{\partial T}{\partial t} + \vec{u}\cdot\nabla T + w  = \nabla^2 T,\nonumber \\
&& \frac{\partial \mu}{\partial t} + \vec{u}\cdot\nabla \mu + \frac{1}{R_0} w  = \tau \nabla^2 \mu,\nonumber \\
&& \nabla \cdot \vec{u} = 0
\label{eq:nondim}
\end{eqnarray}
where ${\rm Pr}$, $\tau$, and $R_0$ were introduced in Section \ref{sec:intro}. 
 To arrive at these equations, we have used the following units:  
\begin{eqnarray}
&&  \left[l\right] = d = \left(\frac{\kappa_T \nu} {\alpha g\left|\frac{dT_0}{dr} - \frac{dT_{\rm ad}}{dr}\right|  }\right)^{1/4} \mbox{ as the unit distance} , \nonumber \\
&& \left[t\right] = \frac{d^2}{\kappa_T} , \quad \, \left[u\right] = \frac{\kappa_T}{d} \mbox{ as the unit time and velocity respectively} , \nonumber \\
&& \left[T\right] =  \left|\frac{dT_0}{dr} - \frac{dT_{\rm ad}}{dr}\right| d \mbox{ as the unit temperature} , \quad \, \left[\mu\right] = \frac{\alpha }{\beta}  \left[T\right]  \mbox{ as the unit composition}
\end{eqnarray}
where $g$ is the local gravity, and all other quantities were defined in Section \ref{sec:intro}. 


In what follows we consider two possible types of domains: 2D domains in the $(x,z)$ plane of size $(250d \times 250d)$, and 3D domains of size $(250d \times L_y \times 250d)$, where $L_y$ shall be varied. We use a fairly large size in the $(x,z)$ plane to allow for the development of large-scale structures in the simulations, should they naturally occur. 
We set the parameters ${\rm Pr}$ and $\tau$ to be both equal to $0.03$, well below unity. The density ratio, on the other hand, is varied to detect and study the existence of different dynamical regimes across the fingering instability range as appropriate. The code used to solve the set of equations (\ref{eq:nondim}) under triply-periodic boundary condition is the pseudo-spectral code described in \citet{Traxleral2011b}. In all cases, we use a resolution of $256$ spectral modes per $250d$ (corresponding to an effective resolution of about 3 grid points per $d$). The same resolution is used in all directions. 

\section{Comparison of 2D and 3D runs at high density ratio}
\label{sec:highR0}

We begin our investigation by considering a ``high'' density ratio case, with $R_0 = 5$. While still significantly below $R_{\rm crit} = 33.3$ (at these parameters), this value is already large enough to be in the sheared regime described by \citet{Radko2010} (see Section \ref{sec:intro}), but small enough to ensure that resolving the buoyancy frequency (which is equal to $\sqrt{\Pr (R_0 - 1)}$ in these units) remains computationally manageable. Indeed, gravity waves are present in this system, and the time step must be small enough to fully resolve their oscillation period. 

\subsection{Results of 2D simulations}
\label{sec:2Dhi}

\begin{figure}
\centerline{\includegraphics[width=0.8\textwidth]{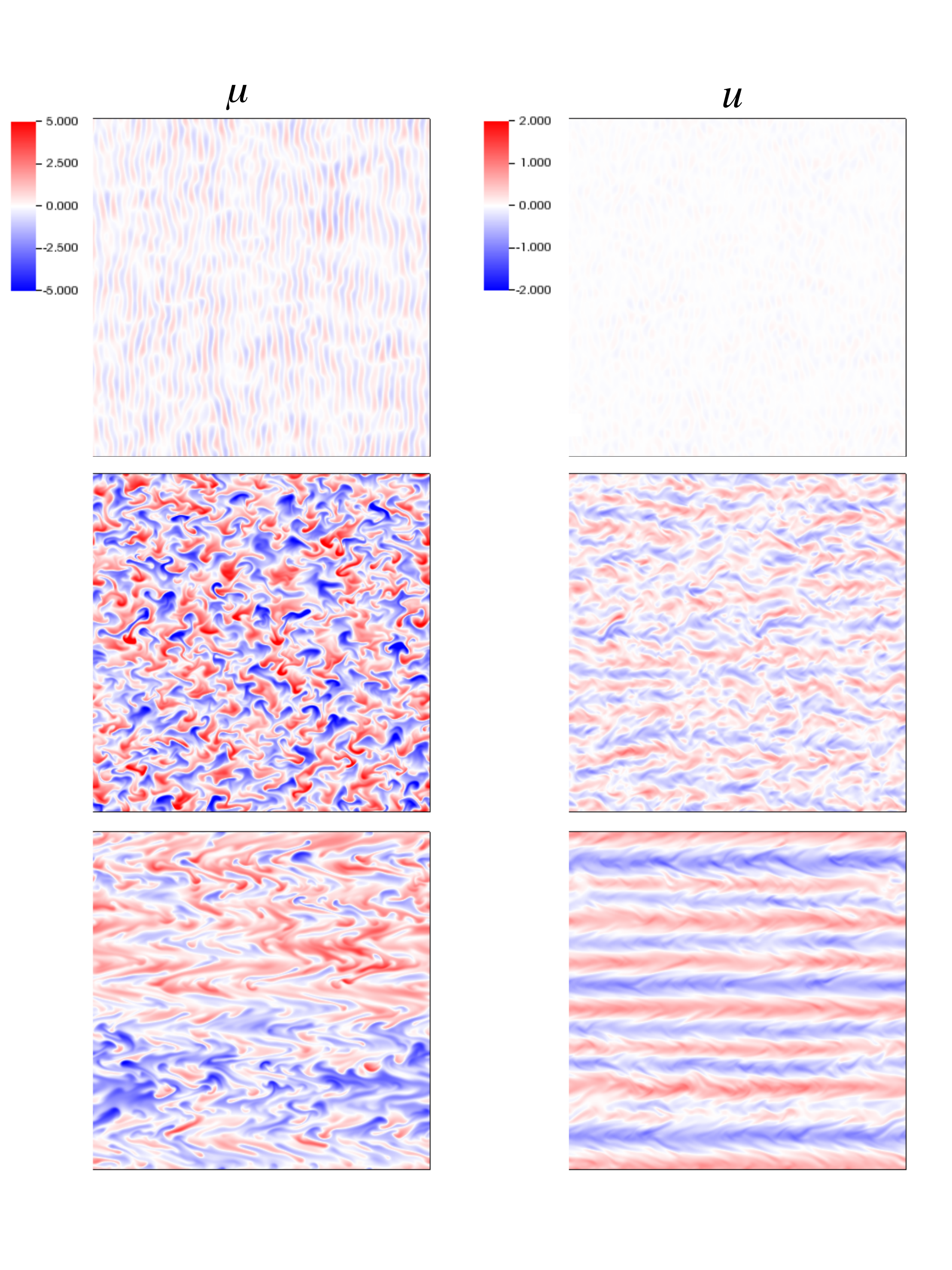}}
\caption{Snapshots of the compositional perturbations (left) and of the horizontal velocity in the $x$ direction (right) at three times (see the evolution also shown in Figure \ref{fig:Nuurms}): near the onset of the fingering instability ($t = 280$, top), early after its saturation ($t = 530$, middle) and at later time once the shear layers have begun to develop ($t =2500$ bottom). Note how the horizontal shear distorts the fingers. }
\label{fig:2DhiR0snaps}
\end{figure}

\begin{figure}
\centerline{\includegraphics[width=0.8\textwidth]{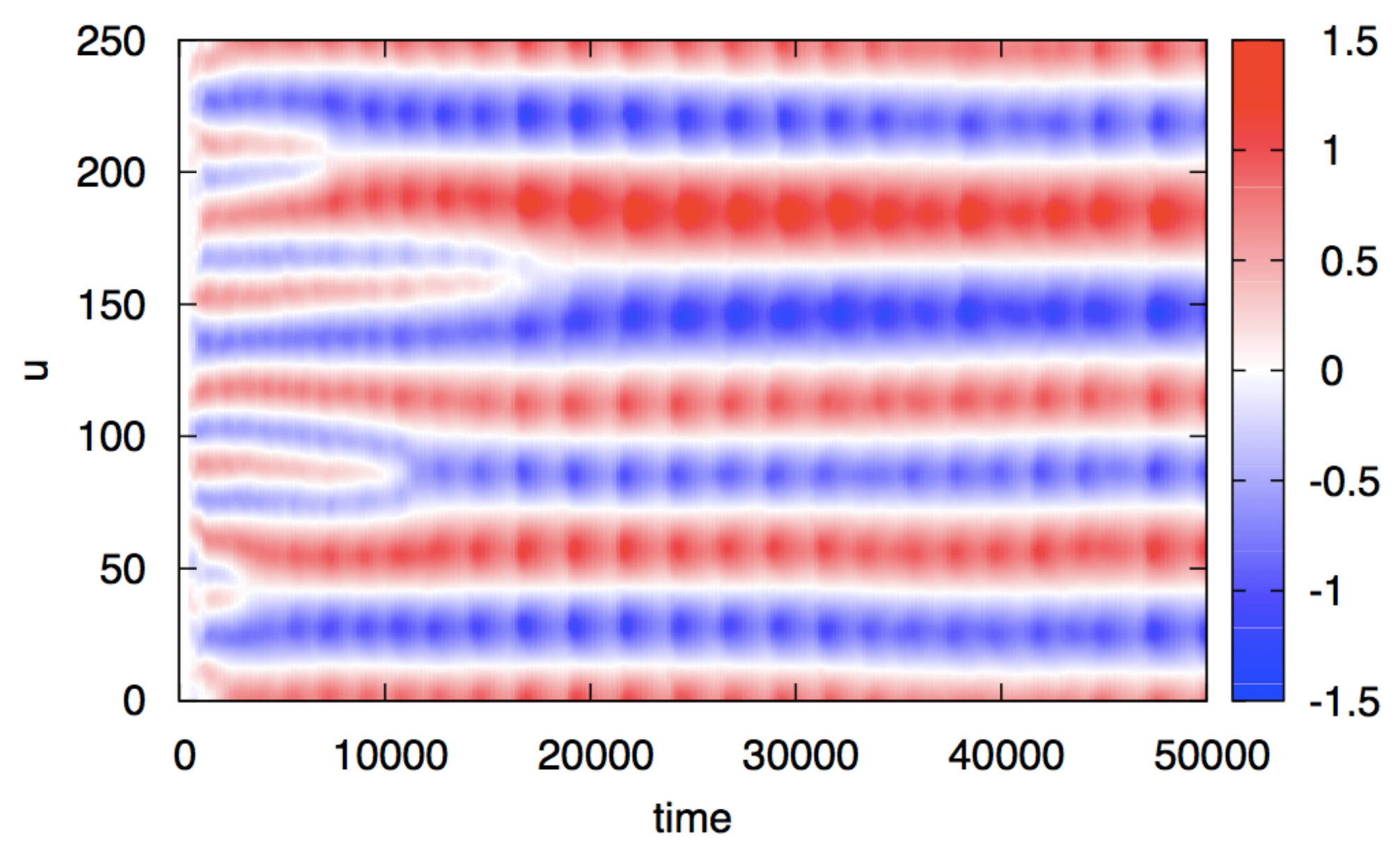}}
\caption{Horizontally-averaged horizontal velocity profile in the 2D simulation of fingering at $R_0 = 5$ as a function of time. The layers rapidly merge down to four, after which a quasi-steady periodic state appears to be reached. The oscillations in the intensity of the shear are clearly visible.}
\label{fig:layermerger}
\end{figure}

Figure \ref{fig:2DhiR0snaps} shows the evolution of the mean molecular weight perturbation $\mu$ and of the horizontal component of the velocity field $u$ in the 2D run, from very early times to much later times. We see, successively, the development of the basic fingering instability (top), its saturation (middle), and finally, the emergence of strong alternating shear layers and their effect on the fingers (bottom). As seen in Figure \ref{fig:layermerger}, the shear layers evolve slowly with time: weak layers appear to be pushed closer to one another, and eventually merge into stronger ones. As discussed in Section 1, the spontaneous emergence of large-scale shear layers in 2D fingering convection at low Prandtl number had already been discovered and studied by \citet{Radko2010}. We therefore confirm his findings here. 


\begin{figure}
\centerline{\includegraphics[width=0.5\textwidth]{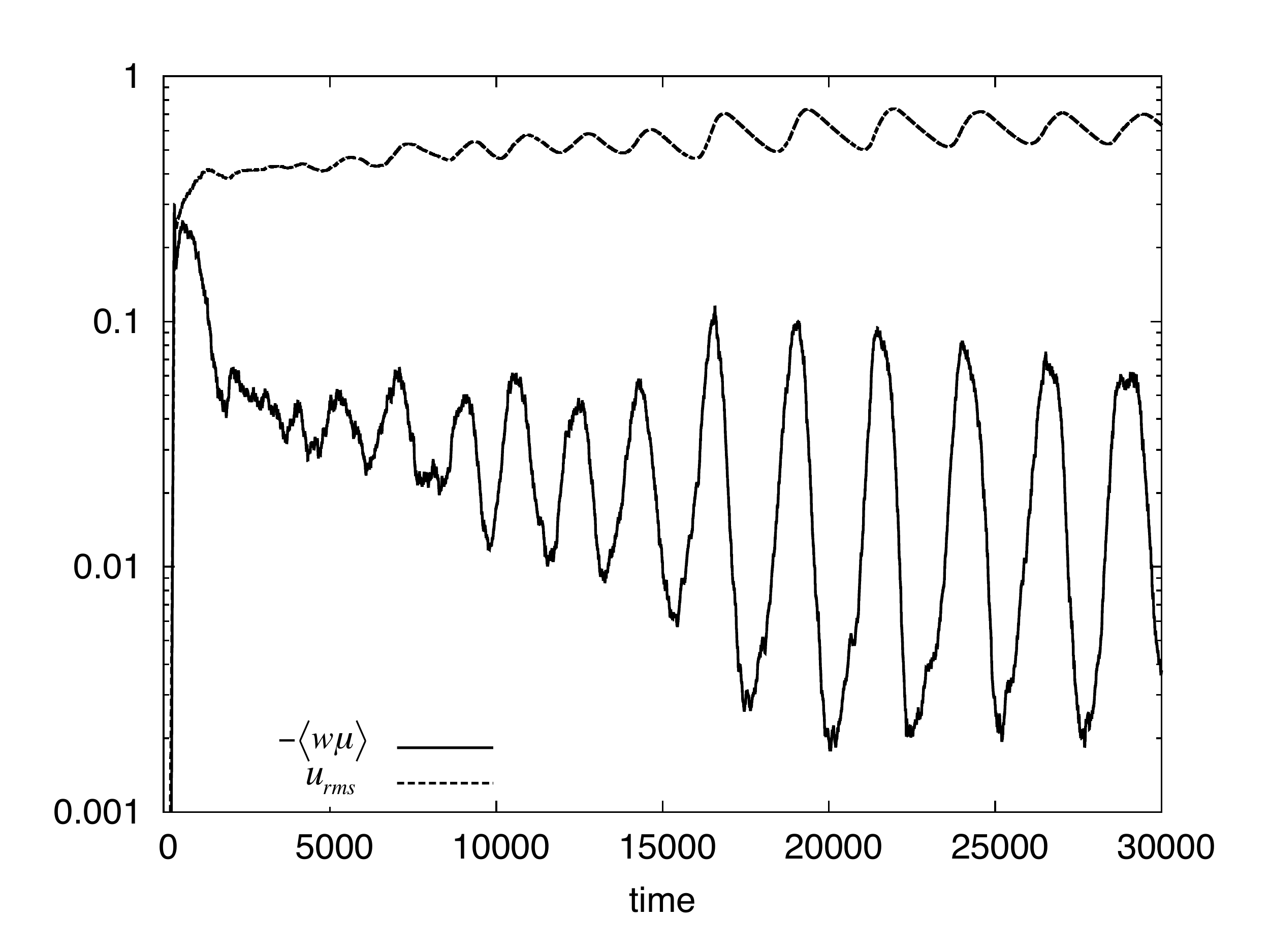}}
\caption{Compositional flux and r.m.s. horizontal velocity in the 2D simulation of fingering at $R_0 = 5$, illustrating the mechanism responsible for the relaxation oscillations.}
\label{fig:Nuurms}
\end{figure}

Interestingly, we further find that the nonlinear interaction of the fingers and the shear drive what appears to be relaxation-oscillations. The latter are most noticeable in Figure \ref{fig:Nuurms}, when viewed in terms of the total compositional flux $\langle w\mu \rangle$ and of the r.m.s. horizontal velocity $u_{\rm rms} = \langle u^2 \rangle^{1/2}$ (where the brackets $\langle \cdot \rangle$ denotes a volume average over the entire domain). The oscillations appear around $t = 5000$, then grow in amplitude as the various shear layers merge. By $t = 15000$, the oscillation pattern becomes very clear: efficient fingering (characterized by a large vertical flux) drives horizontal shear (noticeable in the increase of $u_{\rm rms}$), which first distorts the fingers then eventually completely suppresses fingering convection (hence the drop in the turbulent flux). When this happens, the mechanism that drives the shear dies out, and the shear gradually disappears. The fingers are eventually allowed to grow again, and the cycle repeats. Note that the amplitude and period of the cycle appear to depend on the number of shear layers present. As the simulation progresses and the number of layers decreases through mergers, the cycle lengthens and its amplitude grows, until a quasi-periodic state is reached. 

\subsection{Results of 3D simulations of various $L_y$}
\label{sec:3Dhi}

\begin{figure}
\centerline{\includegraphics[width=\textwidth]{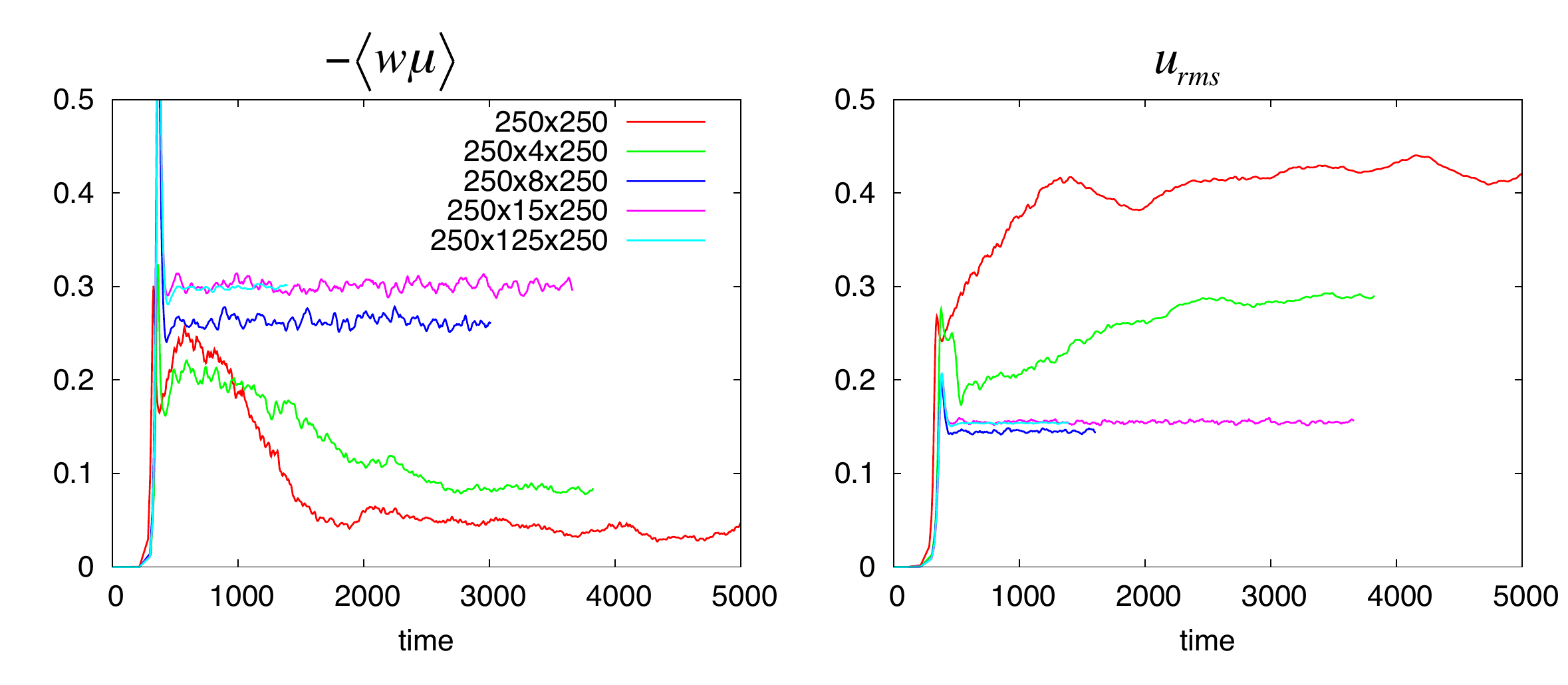}}
\caption{Compositional flux (left) and r.m.s. horizontal velocity (right) for fingering simulations at $R_0 = 5$, of varying domain thicknesses (see legend for detail). Runs with very thin domain behave as if they were 2D, while thick enough domains behave as in a 3D manner that rapidly becomes independent of the domain size. }
\label{fig:2D3DhiR0}
\end{figure}

Even though these fascinating dynamics appear in 2D simulations of fingering convection, it is easy to show that the latter are spurious and do not exist in computational domains that are ``sufficiently 3D". Figure \ref{fig:2D3DhiR0} shows the turbulent compositional flux $\langle w\mu \rangle$ and the r.m.s. horizontal velocity $u_{\rm rms}$ as a function of time for simulations that are perfomed at exactly the same parameters and have $R_0 = 5$, as in Section \ref{sec:2Dhi}, but now in 3D with different domain thicknesses $L_y$. The early stages of the 2D simulation discussed in the previous section are included for comparison.

It is quite clear that the simulation for which $L_y = 4d$ still behaves as if it were 2D, that is, with a turbulent flux that rapidly decreases with time after the initial saturation, and a corresponding r.m.s. horizontal velocity that increases with the generation of shear layers.  By contrast, simulations with $L_y = 15d$ and larger all behave in quantitatively similar ways, achieving the same well-defined quasi-steady state post-saturation, with substantially higher flux and lower horizontal velocities than in 2D and no obvious quasi-periodic oscillations. The run with $L_y = 8d$ appears to have the same character as the wider 3D domains but somewhat underestimates the flux.

The significance of the  cutoff-sizes $L_y = 8d$ and $L_y = 15d$ becomes clearer if one notes that, at the parameter values selected, the typical wavelength of the fastest-growing mode of fingering convection is about 8.5d (one wavelength containing one up-going and one down-going finger). We conclude from this experiment that at low Prandtl number and high density ratio: (1) a domain has to contain at least 1 wavelength of the fastest-growing fingering mode in the third dimension to exhibit dynamics that are fully 3D instead of the spurious shear layers and associated relaxation oscillations that are observed to exist in 2D; (2) a domain has to contain at least 2 wavelengths of the fastest-growing mode to yield quantitatively accurate results for the turbulent fluxes and turbulent intensity of fingering convection. These results are generic:  similar quantitative findings have been obtained for other values of ${\rm Pr}$, $\tau \ll 1$ and for sufficiently large $R_0$. 

\subsection{Can the artificial shear be prevented? }

\begin{figure}[h]
\centerline{\includegraphics[width=0.8\textwidth]{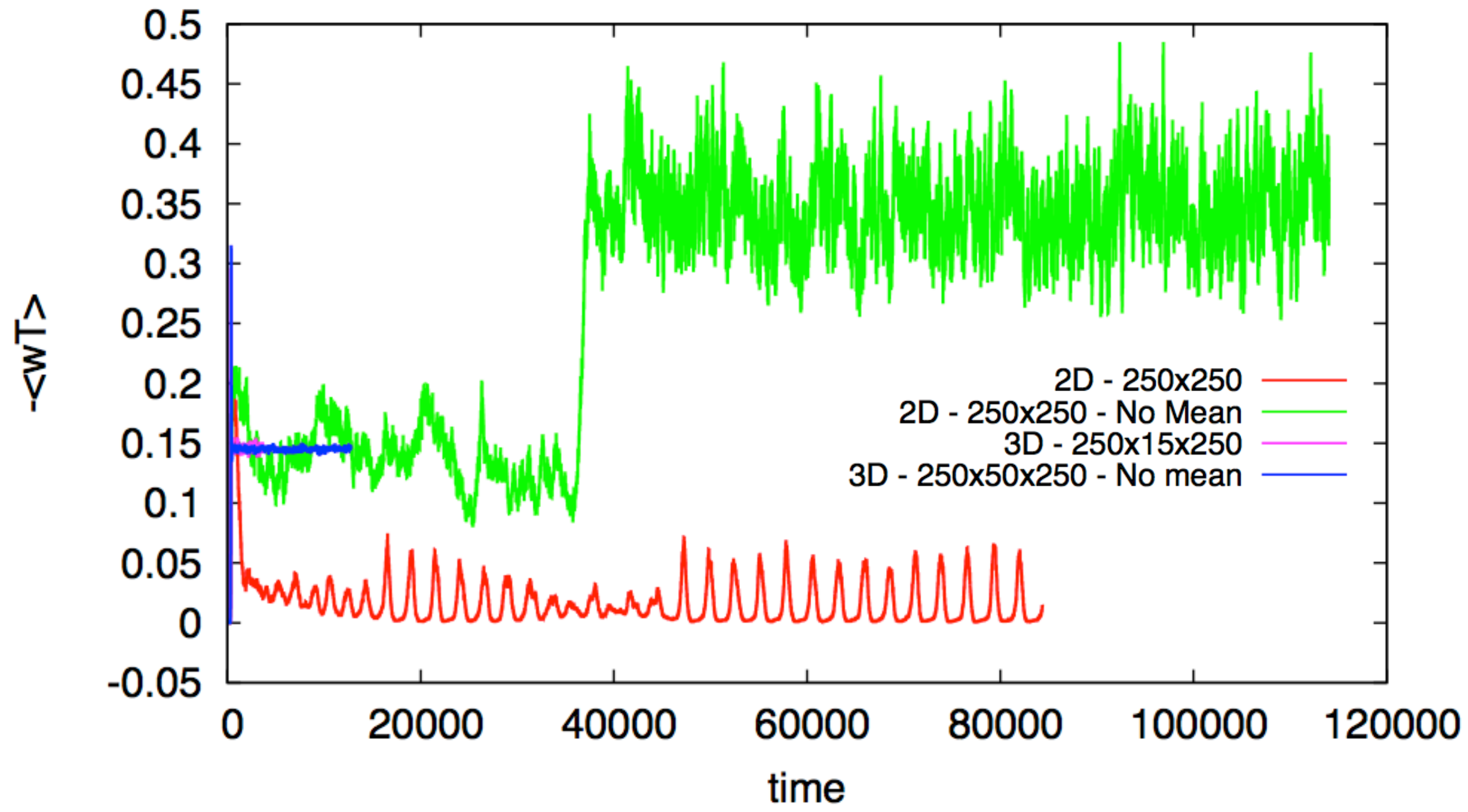}}
\caption{Comparison of time evolution of thermal fluxes for 2D and 3D simulations at $R_0=5$ with and without mean shear flows suppressed.  Fluxes are shown for a purely 2D simulation, a purely 2D simulation with shear suppressed, a narrow 3D simulation that contains 2 finger wavelengths in the narrow direction, and a wider 3D simulation with the shear suppressed.  Clearly, whilst both 3D simulations agree well, both 2D simulations do not reflect the same behavior at all.}
\label{fig:nic1}
\end{figure}

\begin{figure}[h]
\centerline{\includegraphics[width=\textwidth]{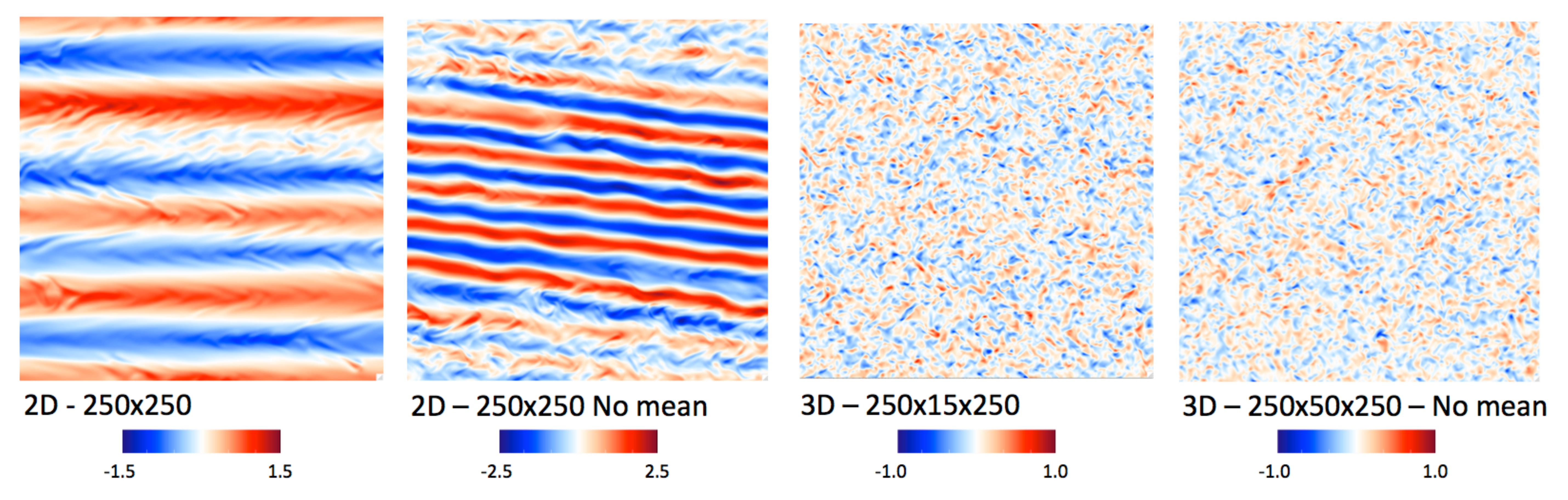}}
\caption{Slices in the $(x,z)$ plane of the horizontal velocity for the simulations shown in Fig. \ref{fig:nic1} at late times in each simulation.  In 2D, strong shear flows are seen, and in 2D with mean shear suppressed, an ``elevator'' solution is observed that is very close to the sheared 2D case but which is periodic.  None of this behavior is seen in 3D, where the suppression of shear makes no difference since significant shear is not present.}
\label{fig:nic2}
\end{figure}

The results of Figure \ref{fig:2D3DhiR0} strongly suggest that the substantial horizontal shear and the associated quenching of the vertical fluxes observed in the 2D case are artificial. However, given the vast savings in computational time between 2D and 3D simulations, it is tempting to wonder if one could still recover acceptable 2D solutions (i.e. solutions whose dynamical behavior and turbulent transport rate are comparable to the 3D solutions) by suppressing the shear. Figures \ref{fig:nic1} and \ref{fig:nic2} demonstrate that this does not work.  Since the code we are using is spectral, it is easy to zero out the mean component of the horizontal flow at every time step, effectively suppressing the shear whilst allowing all other modes to evolve normally.  We performed this experiment on the 2D case at $R_0=5$ discussed in Section \ref{sec:2Dhi}. We compare below these new results to those originally obtained without shear-suppression, and to the narrow 3D case with $L_y = 15d$ discussed in Section \ref{sec:3Dhi} that appeared representative of the true solution.  Finally, in order to ensure that shear-suppression does not artificially affect the dynamics of fingering convection in 3D, we ran a final experiment suppressing the shear in the same manner on another wider 3D case (with domain size $250d\times50d\times250d$) at the same parameters. 

Figure \ref{fig:nic1} shows the thermal flux $\langle wT \rangle$ for each of these cases.   As seen before, the narrow 3D case settles down to a much higher flux than the 2D case, which exhibits relaxation oscillations.  The wider 3D case with the shear suppressed has the same flux as the narrower 3D one, showing that shear does not play a significant role in the 3D simulations.  The 2D case with the shear suppressed, however, does something altogether different from all the other cases.  Although it initially settles into a mode with roughly the right mean flux, the fluctuations about the mean are much larger than in the 3D cases.  Furthermore, at around $t=40000$, the 2D shear-suppressed case jumps to a solution with a much higher flux. 

This rather peculiar behavior is best understood by looking at snapshots of the simulations. Figure \ref{fig:nic2}, which shows slices of the horizontal velocity field, demonstrates the physical differences between these various cases.  In 3D, the solution consists of a homogeneous field of small-scale fingers, whether the shear is removed or not.  In the original 2D case, the presence of strong shear discussed in Section \ref{sec:2Dhi} is obvious.  In 2D with the shear suppressed, it appears that the solution attempts to create a velocity pattern that is as close to the regular 2D case (with strong mean shear) as possible, given the model constraints. After a short transient phase, strong bands of flow appear. They are not exactly horizontal, as that is disallowed, but are slightly angled upwards instead.  To begin with, the specific angles are somewhat random over space. The corresponding vertical flux is significantly larger than in the regular 2D case, and somewhat comparable to the one observed in 3D.  Ultimately though, the flow self-organizes at a single well-defined angle, and acts as an extremely efficient periodic ``escalator" that significantly enhances the vertical transport. Despite being quite interesting, however, these flows remain clearly artificial compared to the 3D cases. We are therefore forced to conclude that there is no easy fix to the problem, and that 3D simulations are the only way to get reliable results.


\section{Comparison of 2D and 3D runs at low density ratio}
\label{sec:lowR0}

\begin{figure}
\centerline{\includegraphics[width=\textwidth]{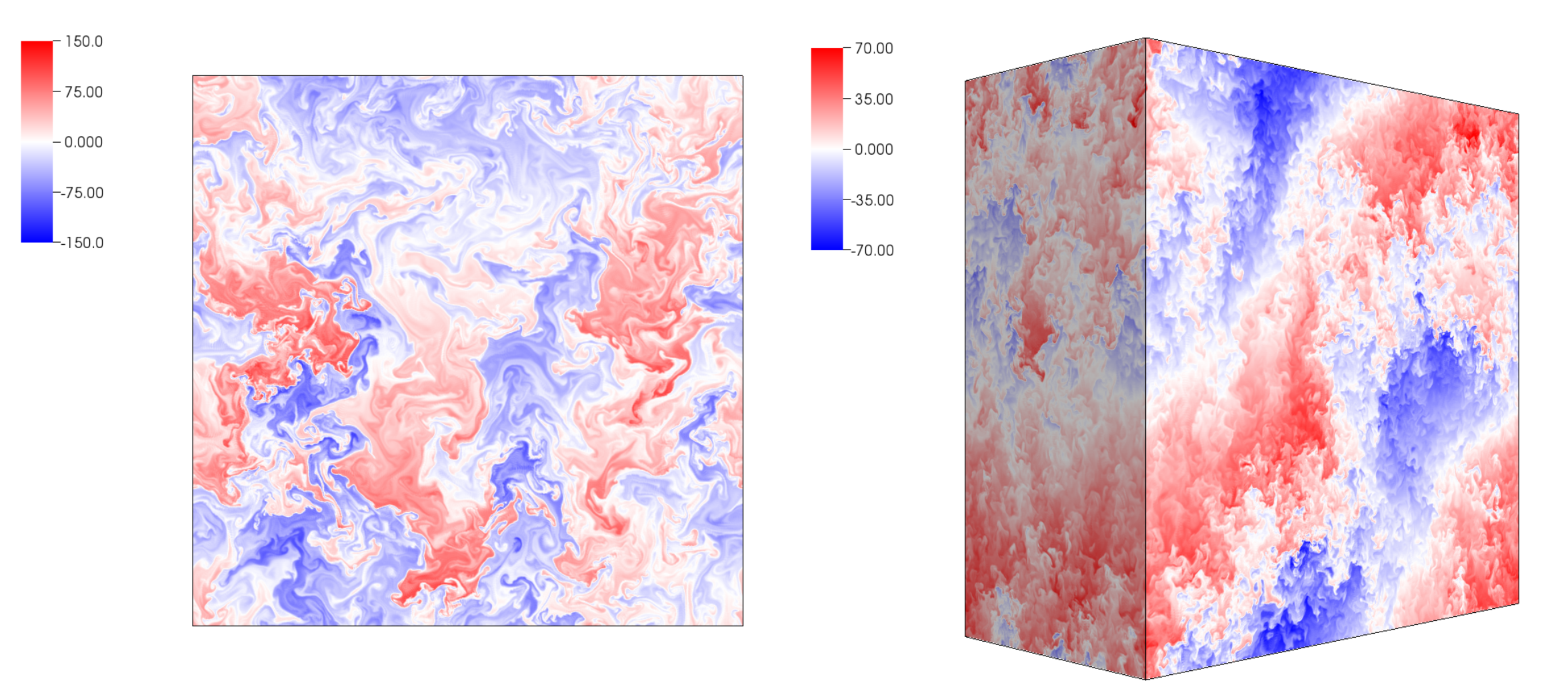}}
\caption{Snapshots of the compositional perturbation for a simulation with $R_0 = 1.11$, in 2D (left) and in a thick 3D domain (right). Note the presence of large-scale gravity waves in the 3D case, absent in the 2D case.  Also note the significant difference in the maximum amplitude of the compositional perturbations in each case.}
\label{fig:2D3DloR0snaps} 
\end{figure}

We now turn to low density ratio systems, and illustrate our findings with a case which has $R_0 = 1/0.9 \simeq 1.11$, and ${\rm Pr} = \tau = 0.03$ as above. 
Figure \ref{fig:2D3DloR0snaps} shows snapshots of a 2D simulation, and of a 3D simulation done in the domain of thickness $L_y = 125d$, both taken around $t = 500$. Again, we see a notable difference between the 2D and the 3D case, but the nature of this difference is clearly not the same as the one discussed in Section \ref{sec:highR0}.
In the 2D case, we see large plumes that are reminiscent of oversized turbulent fingers. It is worth remembering that the typical basic fingering mode width is about 1/30 of the domain size, so the structures seen here (which have a size of about 1/3 of the domain) are much larger than basic fingers. The 3D case, by contrast, exhibits very different dynamics. We note a clear separation of scales between small-scale fingers (which have more or less the same size as the basic instability) and domain-scale gravity waves that modulate the fingering field. 

The emergence of these large-scale gravity waves is well-understood, and can be attributed to the collective instability first discussed by \citet{stern1969cis} in the oceanographic context, later found in simulations of low-Prandtl number fingering convection by \citet{Brownal2013}, and systematically studied by \citet{Medranoal15}. The collective instability is a mean-field instability, excited through a positive feedback loop between the large-scale gradients of temperature and composition, and the respective turbulent fluxes induced by these large-scale gradients\footnote{As discussed by \citet{Radko13}, the collective instability can be interpreted as the turbulent analog of Oscillatory Double-Diffusive Convection, a linear instability that normally takes place in semiconvective regions \citep{Kato66}.}. These large-scale gravity waves do not appear to be excited in 2D, which is somewhat surprising since the collective instability theory does not a priori need to be 3D to operate. 

\begin{figure}
\centerline{\includegraphics[width=0.7\textwidth]{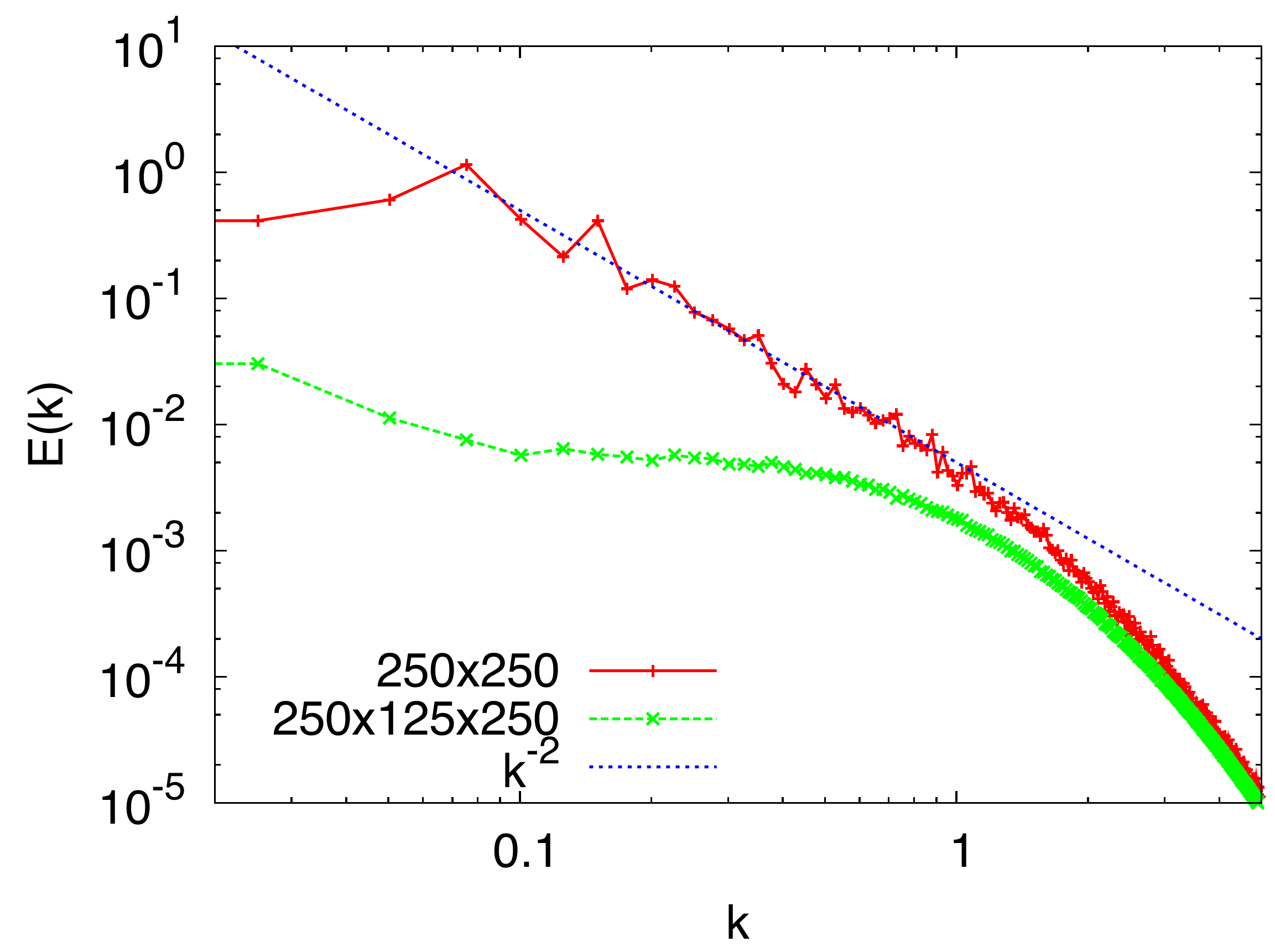}}
\caption{Kinetic energy spectrum as a function of the vertical wavenumber $k$, for the two simulations shown in Figure \ref{fig:2D3DloR0snaps}. The linear instability injection scale is estimated from the horizontal wavenumber $l$ of the fastest-growing mode, and assuming that $ k \sim l$. }
\label{fig:specs}
\end{figure}

A more quantitative way of seeing the scale separation inherent to the collective instability in 3D, and its absence in 2D, is by inspection of the kinetic energy spectrum. This is shown in Figure \ref{fig:specs}  as a function of the vertical wavenumber, for the same simulations and at the same times as the ones shown in Figure \ref{fig:2D3DloR0snaps}. The spectrum of the 2D run clearly peaks around $k  = 0.08 d^{-1}$, which corresponds to a typical wavelength of about $80d$, or in other words, about 10 times the wavelength of the fastest-growing fingering mode. This is more-or-less the scale of the plumes observed in the snapshot. Between that value of $k$ and the energy injection scale ($k \sim 1$) we observe that the kinetic energy spectrum is close to a power law with index $-2$, which may indicate the presence of an inverse energy cascade. The 3D run, by contrast, has a kinetic energy spectrum that peaks at the lowest possible wavenumber $k = 0.025d^{-1}$, which corresponds to a wavelength commensurate with the domain size. This is indeed the scale of the collective instability mode observed in the snapshot. The energy spectrum drops sharply between $k = 0.025d^{-1}$ and $k = 0.1d^{-1}$, then more gently between $k \sim 0.1d^{-1}$ and $k \sim d$ (the energy injection scale).  Beyond $k=d$, both 2D and 3D simulations have the same energy spectrum which is dominated by small-scale diffusive processes, showing that they are basically the same in 2D and 3D. 

\begin{figure}
\centerline{\includegraphics[width=0.8\textwidth]{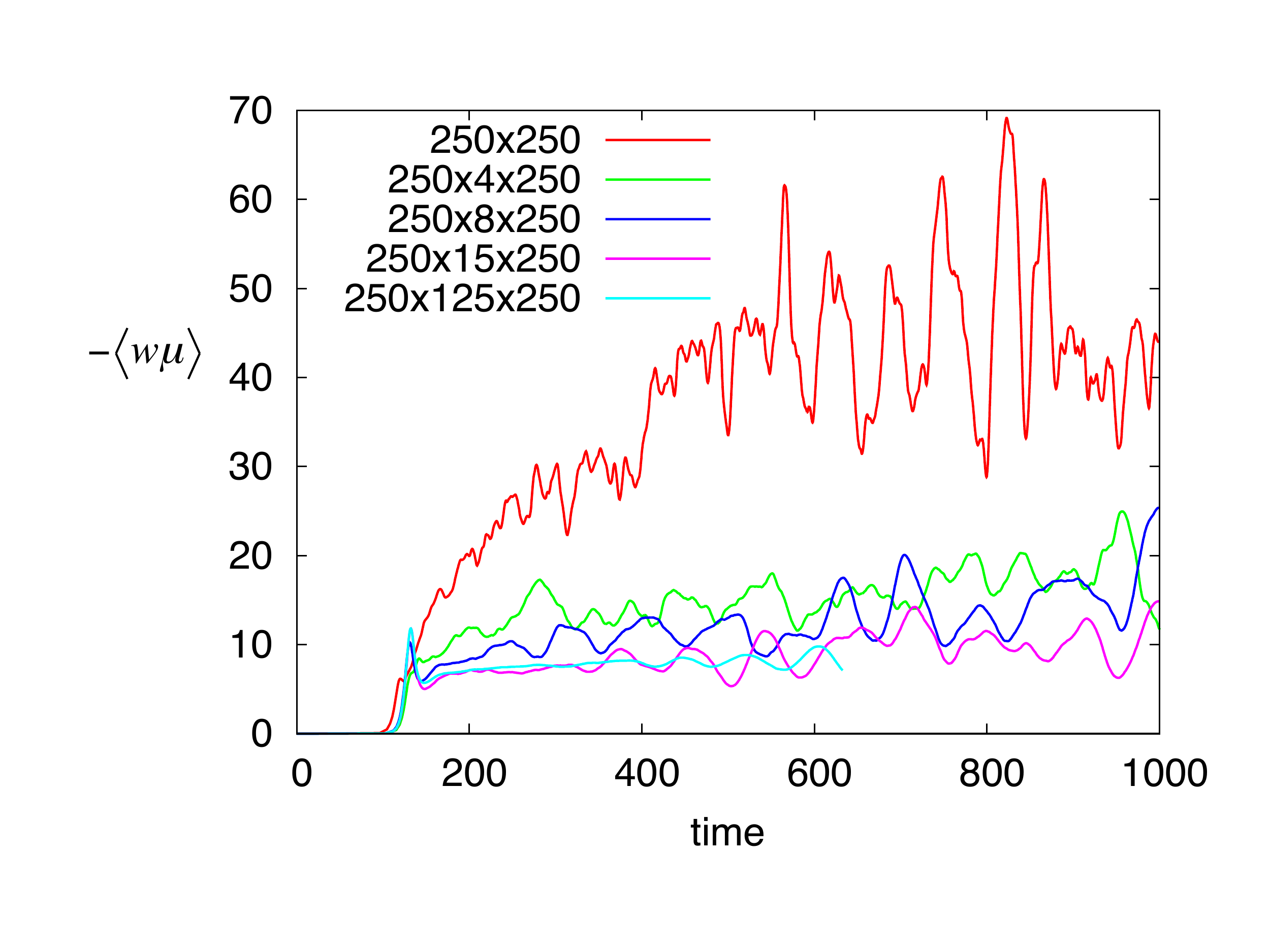}}
\caption{Compositional flux for $R_0 = 1.11$, for various domain thicknesses (see legend for detail). The 2D and 3D runs behave very differently. }
\label{fig:2D3DloR0}
\end{figure}

Finally, Figure \ref{fig:2D3DloR0} shows the turbulent compositional flux for simulations where the thickness of the domain $L_y$ is varied from $0$ to $125d$ as we did in Figure \ref{fig:2D3DhiR0} for the high density ratio case. This time, we see a clear difference in behavior between all the 3D simulations and the 2D simulation. The 2D case shows a continued growth of the efficiency of turbulent transport that only saturates quite late (after $t = 500$) at a fairly high mean value. All the 3D simulations, by contrast, show an early saturation of the primary fingering instability around $t = 150$, then a secondary growth phase associated with the growth of the collective instability, which later also saturates but at a much lower mean value than the 2D case.
By contrast with the high density ratio case described in the previous Section, even the $L_y = 4d$ simulation seems to be qualitatively more compatible with the full 3D case than the 2D case. However, in order to get quantitatively accurate flux measurements just after the saturation of the fingering instability but prior to the onset of the collective instability (i.e. roughly between $t = 150$ and $t = 250$), it is necessary to choose $L_y \ge 15d$, just as in the high density ratio case. In other words, simulations in domains that are at least 2 wavelengths of the fastest-growing mode (which is also about $8.5d$ at this value of $R_0$) are necessary to get adequate estimates of the fingering fluxes at low Prandtl number. 

Of course, even if the fingering dynamics are correctly accounted for, a very narrow domain may no longer be adequate once larger-scale structures appear. In the $L_y = 15d$ domain, the large-scale gravity waves excited by the collective instability are artificially forced to be 2D, for instance. How this affects their saturation amplitude, and their overall behavior, remains to be determined. Similarly, in the work of \citet{Zemskovaal14}, the thin domain used severely restricts the dynamics of the convective layer that eventually forms, to the extent that the results {\it in the convective phase} are not necessarily reliable\footnote{see, for instance, the strong vortices that are clearly visible in their Figure 10, a classic sign of quasi-2D dynamics that doesn't persist in 3D}. In other words, while a thin domain is a good compromise to study homogeneous fingering convection, a full 3D domain with aspect ratio of order unity may remain necessary to properly model any large-scale dynamics that naturally arise. 

\section{Discussion and prospect}
\label{sec:ccl}

In this work we have compared 2D and 3D simulations of fingering convection at low Prandtl number to determine under which circumstance 2D models can be used to approximate  3D systems. We have found that 2D simulations always suffer from some kind of pathology, but that pathology is different for different density ratios. 

For more strongly stratified systems (higher density ratio), 2D simulations produce artificial shear layers that strongly suppress the efficiency of fingering convection, and nonlinearly interact with the latter to produce relaxation oscillations. The emergence of shear at high density ratios in 2D does not come as a surprise, as it was predicted and demonstrated to occur by \citet{Radko2010}. Moreover, a similar phenomenon occurs in 2D Rayleigh-B\'enard convection in vertically-bounded but horizontally-periodic models \citep{Goluskinal14}. The 3D simulations, by contrast, do not show any significant shear. The turbulent fluxes in 3D are significantly higher than in 2D, and are more-or-less steady once the system has reached saturation. 

For more weakly stratified systems (lower density ratio), 2D simulations appear to exhibit an inverse energy cascade which results in the progressive coalescence of fingers into larger and larger ones, a phenomenon that is not seen in 3D. As a result, the fluxes at saturation are significantly higher in 2D than in 3D. Furthermore, this inverse cascade seems to prevent the development of the collective instability, which in 3D drives large-scale gravity waves \citep{Medranoal15}.  

In short, 2D simulations always fail to reproduce the basic dynamics of fingering convection at low Prandtl number and should therefore never be used in this context. The fact that they appear to be adequate for high-Prandtl number studies \citep{stern2001sfu} is, in the light of our work, somewhat surprising. However, this could be due to the fact that the relative importance of the nonlinear terms in the momentum equation is inversely proportional to the Prandtl number (see equation \ref{eq:nondim}), and that in both cases, the offending artificial behavior discovered in 2D is an inherently nonlinear one. A preliminary study reveals that the transition from dynamically-correct to pathologically-sheared 2D fingering convection occurs for Prandtl numbers around 0.5; in other words, for $\Pr \ge 0.5$, 2D simulations provide qualitatively reasonable results, but for $\Pr < 0.5$, 3D simulations are necessary. 

Our findings cast doubt on the 2D fingering flux estimates reported by \citet{denissenkov2010} and \citet{DenissenkovMerryfield2011}. In fact, the simulation snapshots in Figure 4 of \citet{denissenkov2010} are quite similar to our 2D snapshots at $t = 280$ in Figure \ref{fig:2DhiR0snaps}, and clearly show the early stages of the shear development with strongly distorted fingers. Meanwhile, the claim made by \citet{DenissenkovMerryfield2011} that 2D and 3D fluxes at low Prandtl number are very similar could be attributed to the fact that they only integrated their simulations for very short times: we find here that 2D and 3D fluxes are closer to one another before the shear has gained significant amplitude.

We have also found, however, that a fully 3D domain (with equal size in all dimensions) is not necessary to model the correct behavior: a minimum domain width of about 2 wavelengths of the fastest-growing fingering mode is sufficient to eliminate the unwanted behavior, and to yield estimates of the turbulent fluxes that are quantitatively consistent with those obtained in nearly cubic domains. Narrow domains, however, should still be used with caution since they could limit the subsequent development of any large-scale dynamics typically associated with fingering convection (e.g. layer formation and the excitation of large-scale gravity waves). 

Interestingly, while there is a fundamental difference between 2D and 3D fingering simulations, \citet{Mollal15} found that this is not the case for simulations in the oscillatory double-diffusive regime (i.e. in the semiconvective regime) at similarly low Prandtl numbers. There, 2D and 3D simulations behave in qualitatively similar ways, and the fluxes extracted in 2D and 3D are within an order of magnitude of each other for all parameters surveyed. This striking difference between the fingering regime and the oscillatory regime raises a number of questions. Why are they so different, given that the only difference in the governing equations is the sign of the background temperature and compositional gradients? Is the ``two-wavelength" guideline we propose for the selection of a minimal 3D domain size in fingering convection exportable to other instabilities? Generally speaking, can one predict ahead of time and without running simulations whether a given set of equations and boundary conditions, solved for a given set of parameters in 2D, will be a good approximation to the full 3D dynamics? These are clearly difficult mathematical questions, but their answers, if they exist, could provide formal guidelines for the general reliability of 2D and narrow-domain 3D simulations. 

\acknowledgements

P.G. acknowledges funding from NSF AST-1412951. P. G. and N. B. thank Stephan Stellmach for the development of the code used for this work. All simulations were performed on the UCSC Hyades cluster, purchased thanks to an NSF MRI grant.

\addtocontents{toc}{\protect\vspace*{\baselineskip}}




\appendix


\end{document}